\documentclass[conference,10pt]{IEEEtran}
\IEEEoverridecommandlockouts
\usepackage{cite}
\usepackage{amsmath,amssymb,amsfonts}
\usepackage{algorithmic}
\usepackage{graphicx}
\usepackage{textcomp}
\usepackage{xcolor}

\usepackage{booktabs}
\usepackage{multirow}
\usepackage[margin=10pt]{subfig}
\usepackage{soul,color}
\usepackage{listings}
\usepackage[ruled,vlined]{algorithm2e}

\def\BibTeX{{\rm B\kern-.05em{\sc i\kern-.025em b}\kern-.08em
    T\kern-.1667em\lower.7ex\hbox{E}\kern-.125emX}}
    
\DeclareMathOperator*{\argmin}{argmin}
\DeclareMathOperator*{\argmax}{argmax}

\begin{document}

%===========================================================================================================
% CUSTOM COMMANDS
%===========================================================================================================

\newcommand{\moh}[1]{\sethlcolor{orange}\hl{Mohamed: #1}}
\newcommand{\luk}[1]{\sethlcolor{green}\hl{\L ukasz: #1}}
\newcommand{\roy}[1]{\sethlcolor{yellow}\hl{Royson: #1}}

\newcommand{\comment}[1]{}

\newcommand{\til}{{\fontfamily{ptm}\selectfont\texttildelow}}
\newcommand{\xx}{$\times${ }}

%shortcut command for pdf figures
\newcommand{\fig}[3]{\begin{figure}[t]
\centering
\includegraphics[width=2.5in,#2]{#1}
\caption{#3}
\label{#1}
\end{figure}}

%shortcut command for pdf figures
\newcommand{\figvs}[4]{\begin{figure}[!t]
\centering
\includegraphics[width=#1\columnwidth,keepaspectratio,#3]{#2}
\caption{#4}
\label{#2}
\end{figure}}

%shortcut command for pdf figures
\newcommand{\figfull}[4]{\begin{figure*}[!t]
\centering
\includegraphics[width=#1\columnwidth,keepaspectratio,#3]{#2}
\caption{#4}
\label{#2}
\end{figure*}}

\newcommand{\hlc}[2][yellow]{{\sethlcolor{#1} \hl{#2}}}
\newcommand{\addcomment}[2]{\hlc[green]{#1} \hl{#2}}

\comment{
\DeclareMathOperator*{\argmax}{arg\,max}
\DeclareMathOperator*{\argmin}{arg\,min}
\DeclareMathAlphabet{\mymathbb}{U}{BOONDOX-ds}{m}{n}
}

%\comment{
\linespread{0.98}

\addtolength{\textfloatsep}{-10pt}
\addtolength{\dbltextfloatsep}{-10pt}

%Spacing before,after and between figures
\addtolength{\floatsep}{-5pt}
\addtolength{\dblfloatsep}{-5pt}
\addtolength{\abovecaptionskip }{-4pt}
\addtolength{\parskip }{-0.5pt}

%Change caption size to small
\renewcommand{\captionfont}{\small}
\renewcommand{\captionlabelfont}{\small}
%}

%===========================================================================================================

%--------------------------------------------------
\title{Co-optimization of Hardware and DNN Architecture}
\title{FPGA-NAS: Codesign of DNN Hardware and Model Architectures using Reinforcement Learning}
\title{FPGA-NAS: Codesign of CNN and FPGA Accelerator through AutoML}
\title{AutoML Codesign of DNNs and their HW Accelerator on FPGAs}
\title{AutoML Codesign of DNN and FPGA Accelerator}
\title{Co-optimization of Hardware and DNN Architecture for FPGAs}
\title{Best of Both Worlds:\\Codesign of Model Architecture and HW Accelerator for CNNs on FPGAs}
\title{Best of Both Worlds:\\AutoML Codesign of CNN Architecture and FPGA Accelerator}
\title{Best of Both Worlds: AutoML Codesign of CNN Architecture and its FPGA Accelerator for Image Classification}
\title{Best of Both Worlds: AutoML Codesign of CNN Architecture and its FPGA Accelerator}
\title{Best of Both Worlds:\\AutoML Codesign of a CNN and its FPGA Accelerator}
\title{Best of Both Worlds: AutoML Codesign of a CNN and its FPGA Accelerator}
\title{Best of Both Worlds: AutoML Codesign of a CNN and its Hardware Accelerator}
\title{Codesign-NAS: Combined Neural Architecture Search and Design Space Exploration of a CNN and its Hardware Accelerator}
\title{AutoML Codesign of a CNN and its Hardware Accelerator}
\title{Automatic CNN-HW Codesign using Multiobjective Neural Architecture Search}
\title{Codesign-NAS: CNN-HW Codesign using Multiobjective Neural Architecture Search}
\title{Codesign-NAS: AutoML Cooptimization of a CNN and its Hardware Accelerator}
\title{Codesign-NAS: Automatic Co-optimization of a CNN and its Hardware Accelerator}
\title{Codesign-NAS: Automatic CNN/Hardware Co-optimization using Reinforcement Learning}
\title{Codesign-NAS: CNN/Hardware Co-optimization using Reinforcement Learning}
\title{Best of Both Worlds: CNN/Hardware Codesign using Reinforcement Learning}
\title{Best of Both Worlds: AutoML Codesign of a CNN and its Hardware Accelerator}

%--------------------------------------------------
\comment{
\author{Mohamed S. Abdelfattah$^1$, \L ukasz Dudziak$^1$, Thomas Chau$^1$}
\author{Royson Lee$^1$, Hyeji Kim$^1$, Nicholas D. Lane$^{1,2}$}
\email{mohamed1.a@samsung.com}
\affiliation{%
  \institution{$^1$Samsung AI Center, Cambridge, UK}
}
\affiliation{%
  \institution{$^2$University of Oxford, UK}
}
}

\author{\IEEEauthorblockN{Mohamed S. Abdelfattah$^1$, \L ukasz Dudziak$^1$, Thomas Chau$^1$,\\Royson Lee$^1$, Hyeji Kim$^1$, Nicholas D. Lane$^{1,2}$}
\IEEEauthorblockA{\textit{$^1$Samsung AI Center, Cambridge, UK} \\
\textit{$^2$University of Oxford, UK}\\
mohamed1.a@samsung.com}
}

\maketitle

\begin{abstract}
Neural architecture search (NAS) has been very successful at outperforming human-designed convolutional neural networks (CNN) in accuracy, and when hardware information is present, latency as well.
However, NAS-designed CNNs typically have a complicated topology, therefore, it may be difficult to design a custom hardware (HW) accelerator for such CNNs. 
We automate HW-CNN codesign using NAS by including parameters from both the CNN model and the HW accelerator, and we jointly search for the best model-accelerator pair that boosts accuracy and efficiency. 
We call this Codesign-NAS.
In this paper we focus on defining the Codesign-NAS multiobjective optimization problem, demonstrating its effectiveness, and exploring different ways of navigating the codesign search space.
For CIFAR-10 image classification, we enumerate close to 4 billion model-accelerator pairs, and find the Pareto frontier within that large search space.
This allows us to evaluate three different reinforcement-learning-based search strategies.
Finally, compared to ResNet on its most optimal HW accelerator from within our HW design space, we improve on CIFAR-100 classification accuracy by 1.3\% while simultaneously increasing performance/area by 41\% in just \til1000 GPU-hours of running Codesign-NAS.
\end{abstract}

\comment{
\begin{abstract}
Field-programmable gate arrays (FPGAs) have become a popular compute platform for convolutional neural network (CNN) inference; however, the design of a CNN model and its FPGA accelerator has been inherently sequential.
A CNN is first prototyped with no-or-little hardware awareness to attain high accuracy; subsequently, an FPGA accelerator is tuned to that specific CNN to maximize its efficiency.
%Instead, we propose considering both the design of the CNN model and the FPGA accelerator in tandem; furthermore, we automate the codesign of the two for image classification.
Instead, we formulate a neural architecture search (NAS) optimization problem that contains parameters from both the CNN model and the FPGA accelerator, and we jointly search for the best CNN model-accelerator pair that boosts accuracy and efficiency. We call this Codesign-NAS.
In this paper we focus on defining the Codesign-NAS multiobjective optimization problem, demonstrating its effectiveness, and exploring different ways of navigating the codesign search space.
For CIFAR-10 image classification, we enumerate close to 4 billion model-accelerator pairs, and find the Pareto frontier within that large search space.
%Next we propose accelerator innovations and we re-evaluate the Pareto frontier -- we show that our codesign-NAS methodology is an effective way of evaluating new accelerator architectural features, instead of manually tuning an accelerator for a specific CNN.
%Next we propose accelerator innovations that improve the entire Pareto frontier -- we advocate for a design methodology where accelerator features are able to improve the efficiency of a \textit{task} (as represented by a NAS search space) instead of improving the efficiency of a specific CNN.
%Our results show that different accelerator architectures, when paired with a codesigned CNN, can attain both high accuracy and efficiency.
Finally, we compare to ResNet on a highly-tuned accelerator, and show that using codesign, we can improve on CIFAR-100 classification accuracy by 1.8\% while simultaneously increasing performance/area by 41\% in just \til1000 GPU-hours of running Codesign-NAS, thus demonstrating that our automated codesign approach is superior to sequential design of a CNN model and accelerator.
\end{abstract}

\begin{abstract}
We formulate a neural architecture search (NAS) problem that contains parameters from both the convolutional neural network (CNN) model and a hardware accelerator, and we jointly search for the best CNN model-accelerator pair that boosts accuracy and efficiency -- we call this Codesign-NAS.
In this paper we focus on defining the Codesign-NAS multiobjective optimization problem and exploring different ways of navigating the codesign search space.
%For CIFAR-10 image classification, we enumerate close to 4 billion model-accelerator pairs, and find the Pareto frontier within that large search space -- we use this to evaluate different search methodologies for CNN/hardware c codesign.
Finally, we use Codesign-NAS to search for a CNN/hardware pair for a given FPGA platform and we improve upon CIFAR-100 classification accuracy by 1.8\% while simultaneously increasing performance/area by 41\% in just \til1000 GPU-hours of search.
\end{abstract}
}

%\begin{IEEEkeywords}
%component, formatting, style, styling, insert
%\end{IEEEkeywords}
%=================================================================================
\section{Introduction}
%=================================================================================

In the evermore important field of machine learning, there are two trends that are gaining importance and increasing in both volume of academic research and number of industrial deployment.
First, the design of deep neural networks (DNNs) is becoming automated through the general field of automated machine learning (AutoML) and more specifically neural architecture search (NAS).
Instead of manually designing a DNN, NAS automatically searches through hundreds or thousands of models using reinforcement learning (RL), evolutionary algorithms or other approaches, to hone in on the best DNN model.
%Instead of designing a DNN manually, NAS is used to automatically find the best DNN architecture by searching through hundreds or thousands of models, using approaches such as reinforcement learning (RL), evolutionary algorithms, or others.
The second trend is pertaining to the hardware (HW) that runs DNN algorithms.
Instead of using commodity HW, there is an increasing number of custom accelerators, either based on FPGAs~\cite{brainwave, dla} or ASICs~\cite{eyeriss,tpu}.
In this work, we combine NAS for DNN model discovery and HW design space exploration (DSE) to automatically codesign both the DNN model and its HW accelerator.
We expose parameters from both the DNN and the HW thus allowing the bottom-up design of the two parts taking into account accuracy and efficiency of the overall design. 
This, in turn, allows us to tailor the DNN to the HW and vice versa.

\textbf{Related Work.} 
NAS has been successful in discovering DNN models that achieve state-of-the-art accuracy on image classification~\cite{nas_original}, super-resolution~\cite{srnas}, speech recognition~\cite{shrinkml} and machine translation~\cite{translatenas}.
HW-aware NAS adds latency to the reward function so that discovered models optimize both accuracy and inference latency, for example, when running on mobile devices~\cite{mnasnet}. %, proxyless, fnas}.
More recently, reinforcement learning-based \textit{codesign} has been introduced to automate both the discovery of a DNN model and its partitioning across multiple FPGAs based on a theoretical utilization model for each device~\cite{codesign_claim}.
Additionally, the authors in~\cite{codesign_claim2} propose an automatic codesign methodology based on stochastic coordinate descent to refine a DNN search space based on hardware constraints, then codesign the DNN using the refined operations along with a hardware accelerator.

\figvs{1}{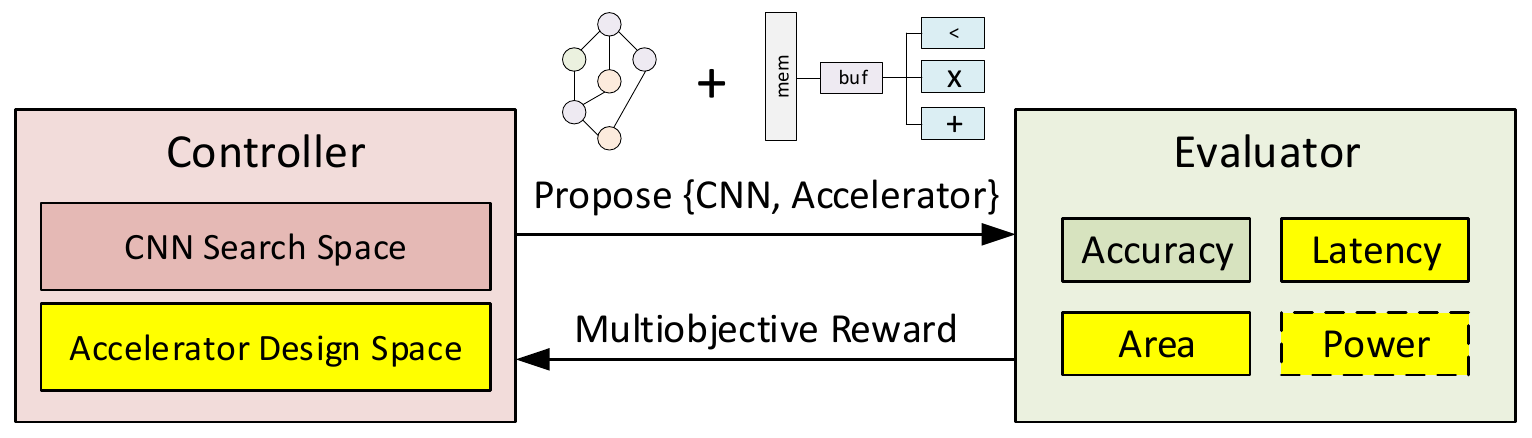}{}{Codesign-NAS system with differences to conventional NAS systems highlighted in yellow.}

\textbf{Contributions.} 
Compared to recent automated codesign literature~\cite{codesign_claim,codesign_claim2}, we use reinforcement learning to automatically codesign CNN and HW architecture, investigate different RL-based search strategies to navigate the codesign search space and demonstrate the efficacy of our search strategies under different scenarios, and finally, we use Codesign-NAS to simultaneously improve accuracy and efficiency of image classification on a popular FPGA platform.
Fig~\ref{codesign_nas} illustrates our system: Codesign-NAS.
A controller selects a CNN architecture from a CNN search space and a HW architecture from an accelerator design space.
Both are sent to the evaluator that implements the CNN on the proposed accelerator to find accuracy and efficiency metrics, such as latency, area and power.
All metrics are then used to create a multiobjective reward that influences the controller to find better CNN-HW pairs.
Our contributions are:
\begin{enumerate}
    \item Propose a general formulation for Codesign-NAS to automatically find CNN-HW pairs using reinforcement learning while optimizing for multiple objectives (area, latency, accuracy).
    \item Enumerate \til 4 billion model-accelerator pairs to study the Pareto-front in a representative codesign search space, and propose three different search strategies to navigate the codesign search space.
    \item Demonstrate the effectiveness of Codesign-NAS by using it for the task of CIFAR-100 image classification. We find CNN-HW pairs that outperform both ResNet~\cite{resnet} and GoogLeNet~\cite{googlenet} even when paired with their most-optimal HW accelerators. (Note that ResNet and GoogLeNet are the best two manually-designed CNNs on our test FPGA platform - we exceed them both on accuracy and HW efficiency).
    %\item Present CNN-HW pairs discovered by Codesign-NAS that improve upon CIFAR-100 image classification on both accuracy and efficiency compared to the optimal HW configurations when running ResNet~\cite{resnet} or GoogLeNet~\cite{googlenet} -- the best two supported CNNs.
\end{enumerate}

%=================================================================================
\section{Approach}
\label{method}
%=================================================================================

In this section we outline our Codesign-NAS system with a focus on the CNN-HW search spaces, the multiobjective optimization (MOO) problem, and the accelerator HW model.

%==========================================
\subsection{Codesign Multiobjective Neural Architecture Search}
%==========================================

NAS focuses on searching for the best parametrization of a predefined model architecture by making a number of \textit{decisions} and evaluating performance of the model when constructed according to the chosen options.
We can define our codesign optimization problem addressed by NAS as:
\begin{equation} \begin{gathered}
\label{opt_search}
    S = O_{nn1} \times  O_{nn2} \times \cdots \times O_{hw1} \times O_{hw2} \times \cdots \\
    s^{\star} = \argmax_{s\in S} \mathcal{E}(s)
\end{gathered} \end{equation}
%where $n$ is the number of decisions (hyperparameters of the model), $O_i$ is the set of available options for the $i$-th decision and $\mathcal{E}$ is the evaluation function used to obtain a model's performance.
% moh: @lukasz, I changed this a bit, let me know what you think
where $O_{nn}$ is a DNN option, such as the selection of an operation, and $O_{hw}$ is a hardware option such as the size of a buffer.
$\mathcal{E}$ is the evaluation function to find the performance of a search point $s$; that is, finding the accuracy and efficiency metrics of running the discovered DNN on the discovered HW.

Since enumerating all points from $S$ is often infeasible in practice due to their large number and time-consuming evaluation, the main challenge of NAS is to produce as good an approximation of $s^{\star}$ as possible while being allowed to evaluate only a limited number of architectures.
Therefore, for a sequence of $T$ explored architectures (search steps): $\tau(T)=(s_1, s_2, \cdots, s_T)$, Eq.~\ref{opt_search} takes the following form:
\begin{equation} \begin{gathered}
\label{approx_search}
    s^{\star} \approx s^{\ast} = \argmax_{s\in \tau(T)} \mathcal{E}(s)
\end{gathered} \end{equation}
and the main focus is to guarantee that the search is able to explore points which optimize $\mathcal{E}$ as $T$ increases.
%We distinguish between Equations~\ref{opt_search} and~\ref{approx_search} here because we are able to perform both in this work and compare them against each other.
%In Section~\ref{analysis}, we use a CNN search space that contains precomputed accuracy, thus allowing us to find exactly the Pareto-optimal points within the space (Equation~\ref{opt_search}), and then compare them to the points found by our search algorithms (Equation~\ref{approx_search}) under different search strategies and optimization targets.

In this work, we use a probabilistic, trainable policy $\pi_{\theta}$ to guide the search as proposed in prior work~\cite{nas_original}.
At each search step $t$ the policy (implemented as a single LSTM cell followed by a linear layer as in~\cite{nas_original}) is first sampled in order to get a structure sequence $s_t$ and later updated using REINFORCE and stochastic gradient descent: $\nabla_{\theta} \pi_{\theta}(s_t)\mathcal{E}(s_t)$.

We consider the following three \textit{quality metrics} when assessing a DNN-accelerator pair: accuracy of the DNN, area of the accelerator and latency of the DNN when run on the accelerator.
In order to link the three objectives to the original optimization problem from Eq.~\ref{opt_search}, we consider a mixture of two standard approaches: we first limit the set of points in $S$ by providing a set of thresholds for all/some of the metrics and filter out points with at least one of them being above/below the threshold ($\epsilon$-constraint~\cite{moo_book}); we then take the weighted sum of the remaining ones~\cite{moo_book}.
This is expressed by the following \textit{reward} function $\mathcal{R}$:
\begin{equation}\begin{gathered}\label{eq:generic_moo}
    \mathcal{R}: \{ \mathbf{m} \ |\ \mathbf{m} \in \mathbb{R}^n \land \forall_i [ \mathbf{m}_i \ge \mathbf{th}_i ] \} \rightarrow \mathbb{R}\\
    \mathcal{R}(\mathbf{m}) = \mathbf{w} \cdot \mathcal{N}(\mathbf{m})
\end{gathered}\end{equation}
where $\mathcal{N}$ is a linear element-wise normalization function which maps values from the range $(x_{\text{min}}, x_{\text{max}})$ to $(0,1)$, $\mathbf{m}$ is a vector of metrics, $\mathbf{th}$ is a vector of thresholds and $\mathbf{w}$ is a vector of weights.
The evaluation function, which is the subject of optimization according to Eq.~\ref{opt_search}, is defined as:
\begin{equation} \begin{gathered}
    \mathcal{E}(s) = \mathcal{R}\big(-\text{area}(s), -\text{lat}(s), \text{acc}(s)\big)
\end{gathered} \end{equation}

If a search point does not meet specified constraints, a \textit{punishment} function $\mathcal{R}_v$ (with opposite sign to the reward) is used as feedback for the RL controller to deter it from searching for similarly \textit{bad} points.

%==========================================
\subsection{Codesign Search Space}
%==========================================

%==========================================
\subsubsection{CNN Search Space}
%==========================================

%
\figvs{0.9}{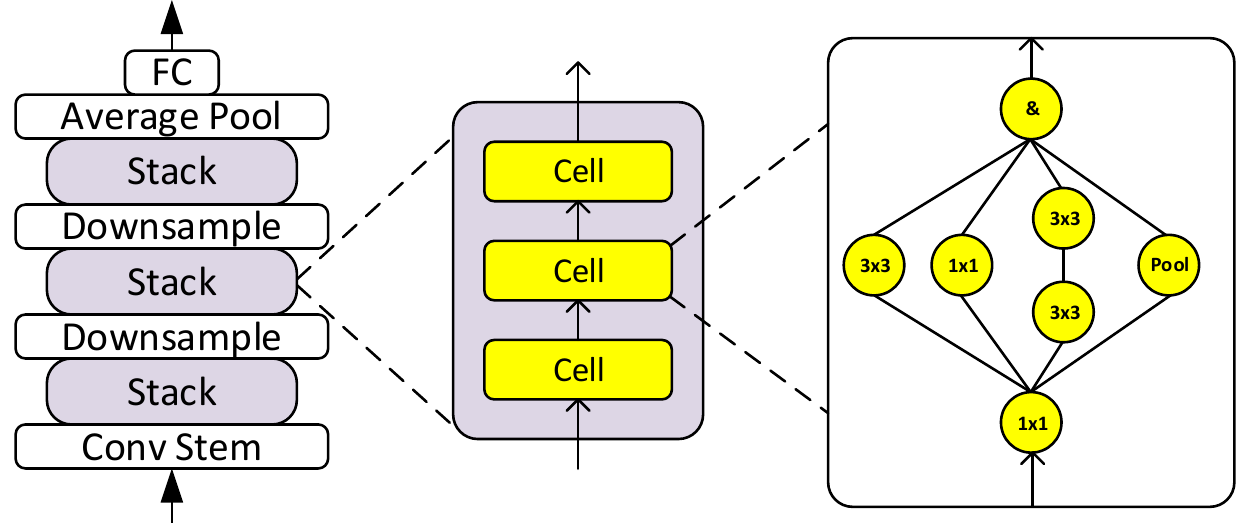}{}{NASBench~\cite{nasbench101} CNN architecture.}
%
%
%\figvs{0.9}{lstm_controller}{trim = 0 0.5cm 0 0 }{Our codesign LSTM controller samples both a CNN and an FPGA accelerator.}
%

NASBench~\cite{nasbench101} provides a CNN search space for image classification including evaluation statistics such as accuracy and training time on the CIFAR-10 dataset for \til 423 thousand unique CNNs.
Fig.~\ref{nasbench_ss} shows the structure of the CNNs within NASBench.
The only varying part of each model is the inner-most design of a single cell, which is then repeated three times to form a stack.
At most 7 operations and 9 connections are allowed per cell, and addition/concatenation operations are inserted automatically according to a set of rules -- for more information, please refer to NASBench~\cite{nasbench101}.
In Section~\ref{analysis}, we use the NASBench database of precomputed accuracy to find exactly the Pareto-optimal points within the codesign space (Equation~\ref{opt_search}), we then compare them to the points found by our search (Equation~\ref{approx_search}) under different search strategies and optimization targets.

%==========================================
\subsubsection{Accelerator Design Space}
%==========================================

%
\figvs{0.9}{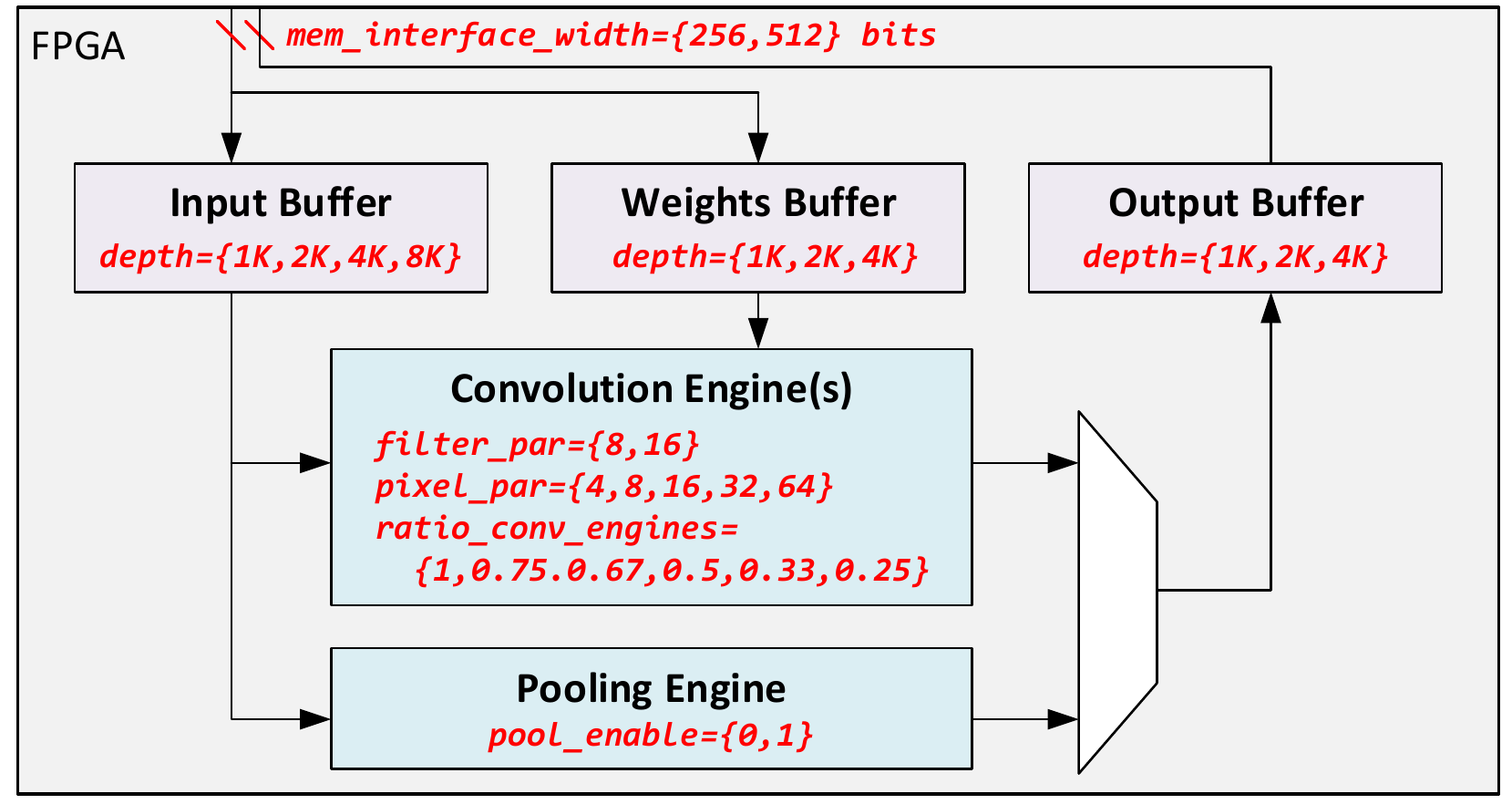}{}{CHaiDNN~\cite{chai} FPGA accelerator with configurable parameters highlighted in red.}

\comment{
\begin{small}
\begin{table}[th]
    \centering
    \caption{ChaiDNN design space parameter options.}
    \begin{tabular}{cc}
         \toprule
          Parameter            & Options              \\
         \midrule
          filter\_par            & \{8, 16\}              \\
          pixel\_par             & \{4, 8, 16, 32, 64\}              \\
          input\_buffer\_depth   & \{1K, 2K, 4K, 8K\}              \\
          weights\_buffer\_depth & \{1K, 2K, 4K\}                  \\
          output\_buffer\_depth  & \{1K, 2K, 4K\}                  \\
          mem\_interface\_width  & \{256, 512\}                  \\
          pool\_en               & \{false, true\}                 \\
          ratio\_conv\_engines$^*$   & \{1, 0.75, 0.67, 0.5, 0.33, 0.25\}              \\
         \bottomrule
         \multicolumn{2}{c}{$^*$ Not present in original ChaiDNN~\cite{chai},}
    \end{tabular}
    \label{chai_params}
\end{table}
\end{small}
}

We base our work on CHaiDNN -- a library for acceleration of CNNs on System-on-chip FPGAs~\cite{chai}.
%It consists of a CPU acting as the control unit and scheduler, an external memory, and an FPGA accelerator.
The FPGA accelerator supports convolution and pooling operations while the unsupported layers run on the CPU.
The accelerator is configurable to maximize the hardware efficiency on different CNNs.
Fig~\ref{chaidnn} shows the accelerator parameters and their valid values. 
%Table~\ref{chai_params} lists 
The available parameters are fairly standard ones in custom hardware accelerators, configuring things like buffer depths, external memory interface width, and the amount of parallelism in the filter and pixel dimensions.
We add the parameter $ratio\_conv\_engines$ to CHaiDNN.
In its default configuration, CHaiDNN sets $ratio\_conv\_engines = 1$, which means that a single general convolution engine runs any type of convolution.
When $ratio\_conv\_engines$ is set to any number below 1, there are two convolution engines - one of them specialized for 3x3 filters, and the other for 1x1 filters -- the ratio determines the number of DSPs assigned to each convolution engine.
These parameters form 8640 different combinations of valid CHaiDNN accelerators, and is representative of a relatively simple hardware design space.
While we use this search space to implement our system and prove its effectiveness, we believe that much more parameter-rich hardware design space can be equally leveraged using our methodology, and would be expected to achieve larger improvements.

%------------------------------
\subsection{Accelerator Modelling}
%------------------------------

%To enable fast and distributed execution with Codesign-NAS, it is essential to have a fast evaluator for the accelerator to find efficiency metrics such as area and latency.

To avoid time-consuming compilations/runs on the FPGA, we create area/latency models to use with Codesign-NAS.

%However, the runtime of NAS is prohibitive if we compile every accelerator architecture online to measure area and latency. 
%Furthermore, running multiple CNN models simultaneously with NAS requires 100s of FPGA boards.
%In this section, we describe how we model CHaiDNN for fast evaluation of area and latency in Codesign-NAS.

%------------------------------
\subsubsection{Area model}
%------------------------------

%\figvs{1}{validation}{trim = 1cm 1cm 2cm 1.5cm}{Estimated and measured area and latency of CHaiDNN at different parallelism. Other parameters: $pool\_enable=1$, $mem\_interface\_width=512 bits$, $buffer\_depths=(8K,2K,2K)$.}

%
\begin{table}[!t]
\centering
    \begin{footnotesize}
    \caption{Estimated FPGA block area for Zync Ultrascale+.}
    \begin{tabular}{lcc}
    \toprule
    Resource & Relative Area (CLB) & Tile Area ($mm^2$)\\
    \midrule
		CLB                   & 1            & 0.0044\\
		BRAM - 36 Kbit        & 6            & 0.026\\
		DSP                   & 10           & 0.044\\
	\midrule
		Total                 & 64,922        & 286 \\
    \bottomrule
    \end{tabular}
    \label{zync_area}
    \end{footnotesize}
\end{table}

We divide the accelerator into its different components: convolution engine, buffers, pooling engine, memory interface, and we create area models based on the utilization of configurable logic blocks (CLB), digital signal processors (DSP) and block RAM (BRAM).
For each component, we break it down further into simpler subcomponents -- for example, a sliding window buffer within the convolution engine that is parameterized with $pixel\_par$ and $filter\_par$ would be modeled with an equation that takes these 2 variables as input.
%We model the number of BRAMs with equation that takes those 2 variables as input.
We verified our area model against 10 full FPGA compilations with different parameters and our model had on average 1.6\% error -- we felt this was adequate for area estimation.
Based on the FPGA resource utilization, we estimate the accelerator size in $mm^2$ such that area is quantified by a single number -- silicon area\footnote{Silicon area is not available for Zync Ultrascale+ devices so we use area for similar devices~\cite{fpt12} and account for the process node (20nm vs. 40nm) and the different block properties (8 LUTs per CLB instead of 10, and 36 Kbit per BRAM instead of 9 Kbit).} according to Table~\ref{zync_area}. 

%------------------------------
\subsubsection{Latency model}
%------------------------------

The latency model consists of two parts: 1) latency lookup table of operations and 2) scheduler.
%Our CNN search space contains just 3 options for operations: 3x3, 1x1 convolutions and 3x3 maximum pooling.
In our CNN search space, there are 85 unique variations of convolutions, pooling and element-wise operations (different input/filter sizes etc.).
We run each operation on the FPGA accelerator with different parameters and measure latency which we then store in a lookup table. 
The scheduler assigns operations to the parallel compute units greedily and calculates the total latency of the CNN model using the lookup table.
%Note that we scale the performance of dual convolution engines based on the parameter $ratio\_conv\_engines$.
%
%
To validate the latency model, we pick the CNN model from NASBench with the GoogLeNet cell, and we run it on 10 different accelerator variants with different parameterizations.
Our latency model is 85\% accurate on this validation set - there is room for improving our per-layer latency model, but we leave that to future work.
%Fig.~\ref{validation} shows that our latency model closely tracks the measured points for different accelerator parallelism parameters.
%Even though we measure latency per-layer, we still get fairly accurate estimates for end-to-end model latency.

%=================================================================================
\section{Search Investigation on NASBench-101}
\label{analysis}
%=================================================================================

In this section, we analyze the codesign search space using the NASBench dataset and our accelerator latency/area model.
The CNN accuracy in NASBench is precomputed and stored in a database, and our FPGA accelerator model runs quickly on a desktop computer.
This allows us to enumerate the entire search space, consisting of 3.7 billion data points, and find the Pareto-optimal points within that space.
Finally, we investigate how to automatically navigate the codesign search space using our RL-based methodology described in Section \ref{method}.
In that context, we evaluate three search strategies in terms of proximity of discovered points to the Pareto-optimal ones and the search convergence speed.

%-------------------------
\subsection{Pareto-Optimal Points}
%-------------------------

To understand the \textit{good} points in the codesign search space, we look for Pareto-optimal~\cite{moo_book} points within the 3.7 billion model-accelerator pairs.
This is done iteratively by filtering dominated points from the search space.
The remaining (non-dominated) points are better in at least one of our evaluation metrics (area, latency or accuracy) w.r.t. any other point.
For our search space, there were only 3096 Pareto-optimal CNN-HW pairs -- these are illustrated in Fig.~\ref{pareto_arch}.

\figvs{1}{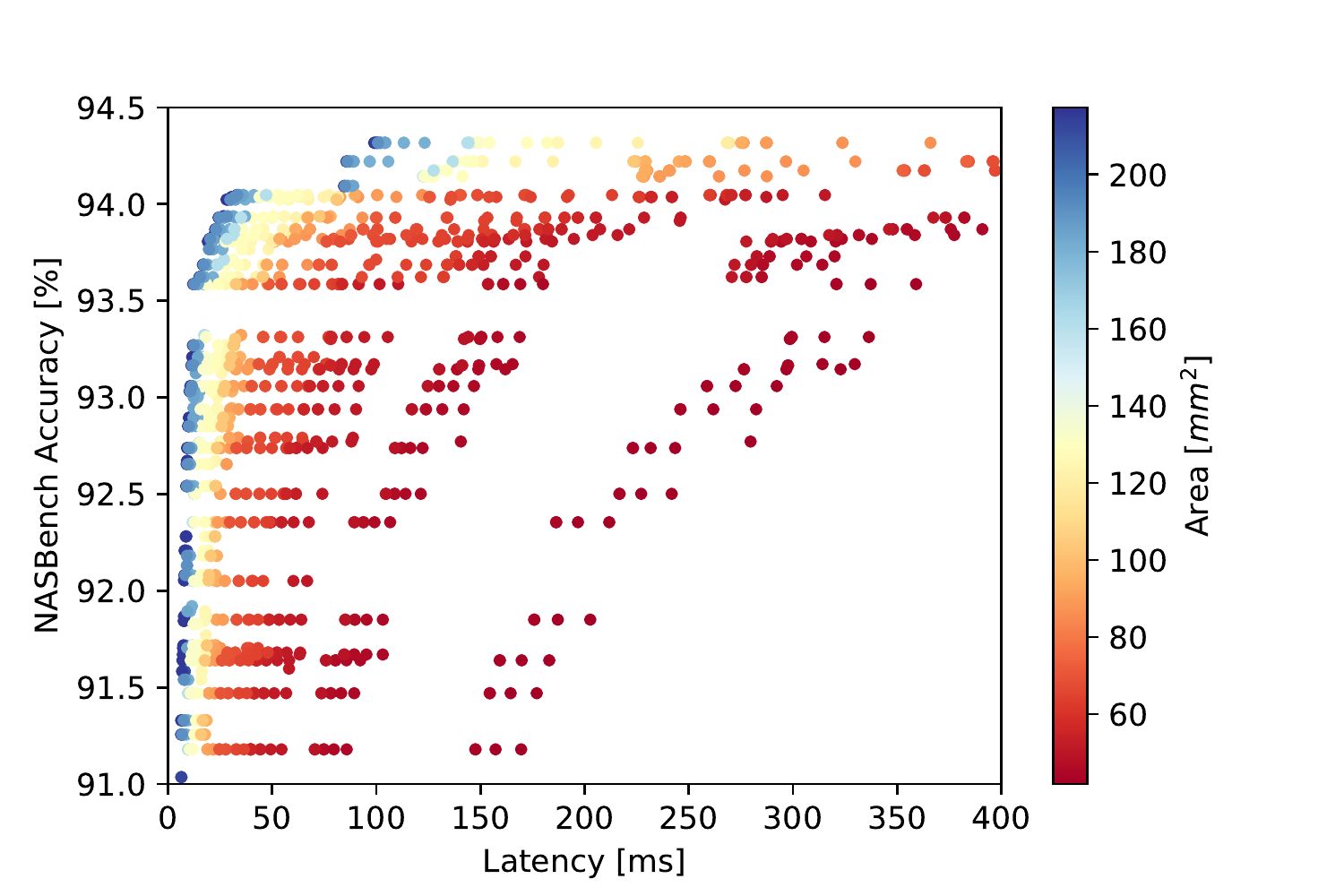}{trim=0 0cm 0 1cm}{Pareto-optimal points in the codesign search space.}
\begin{figure*}[t]
\centering
\subfloat[Unconstrained]{
   \includegraphics[width=0.67\columnwidth,keepaspectratio, trim=0 2mmm 0 11mm]{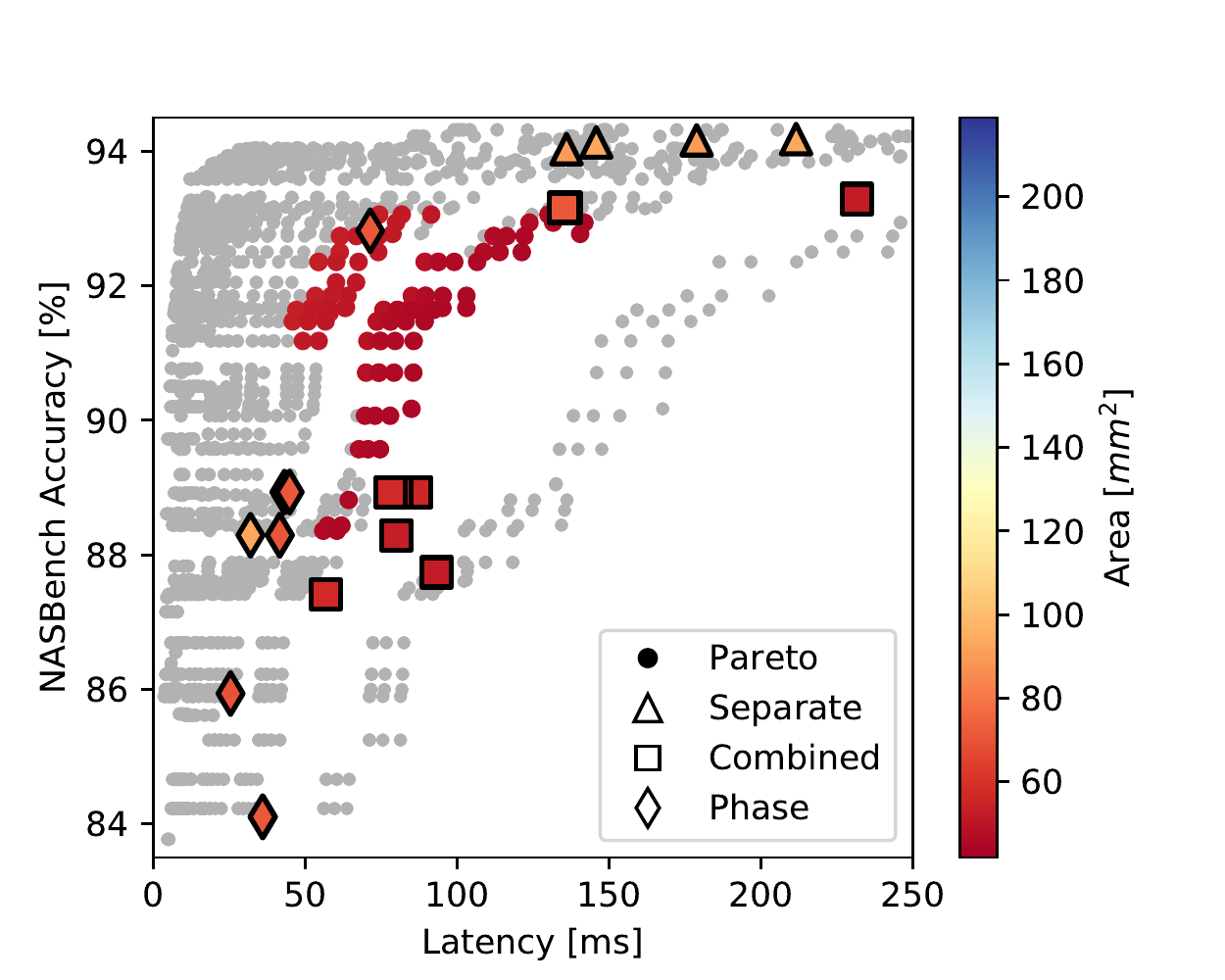}
   \label{e0}
 }
\subfloat[1 Constraint]{
   \includegraphics[width=0.67\columnwidth,keepaspectratio, trim=0 2mmm 0 11mm]{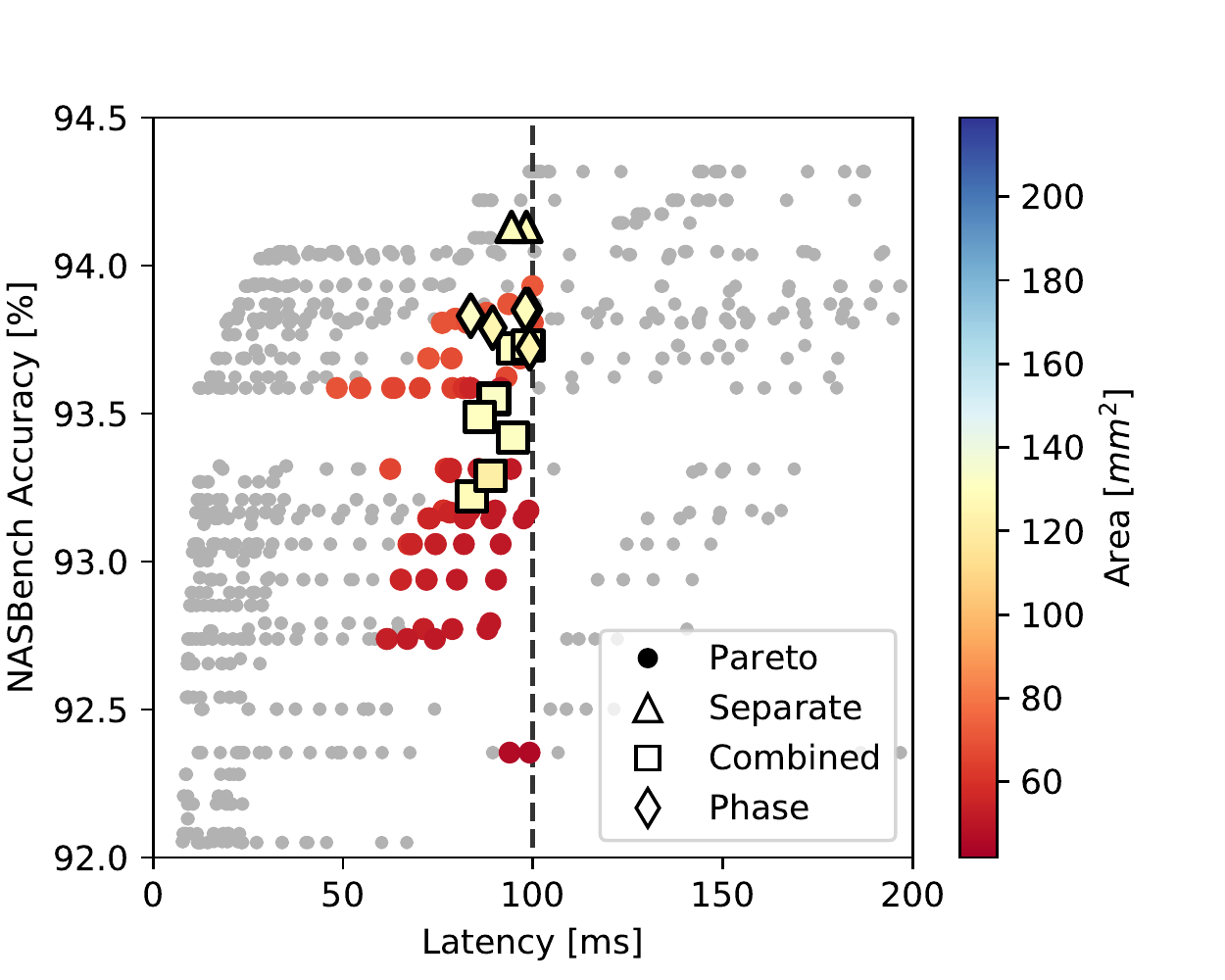}
   \label{e1}
 }
\subfloat[2 Constraints]{
   \includegraphics[width=0.67\columnwidth,keepaspectratio, trim=0 2mmm 0 11mm]{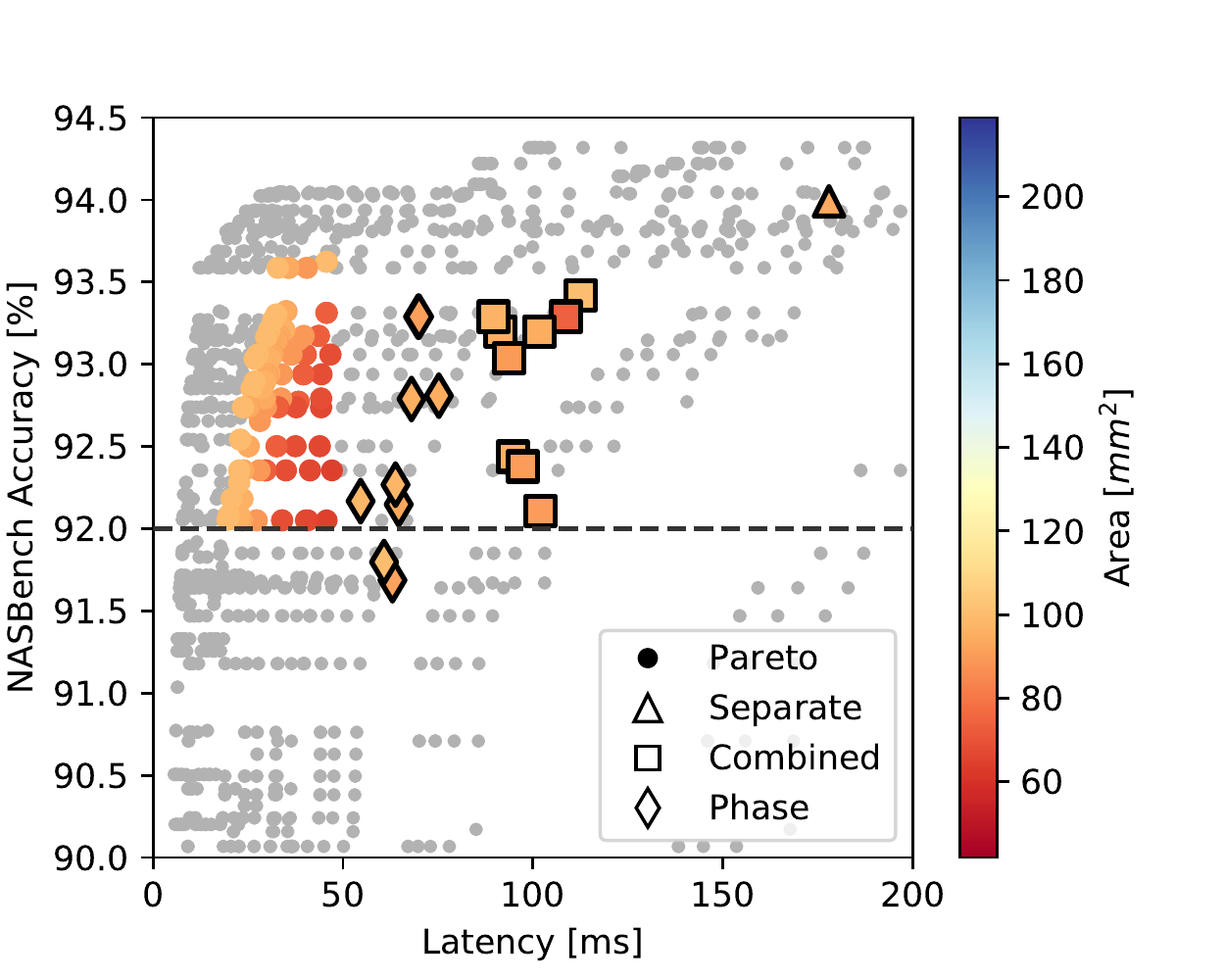}
   \label{e2}
 }
\caption{Top search results compared to the top 100 Pareto-optimal points that maximize each experiment's reward function.}
\label{searches}
\end{figure*}
\figvs{1.05}{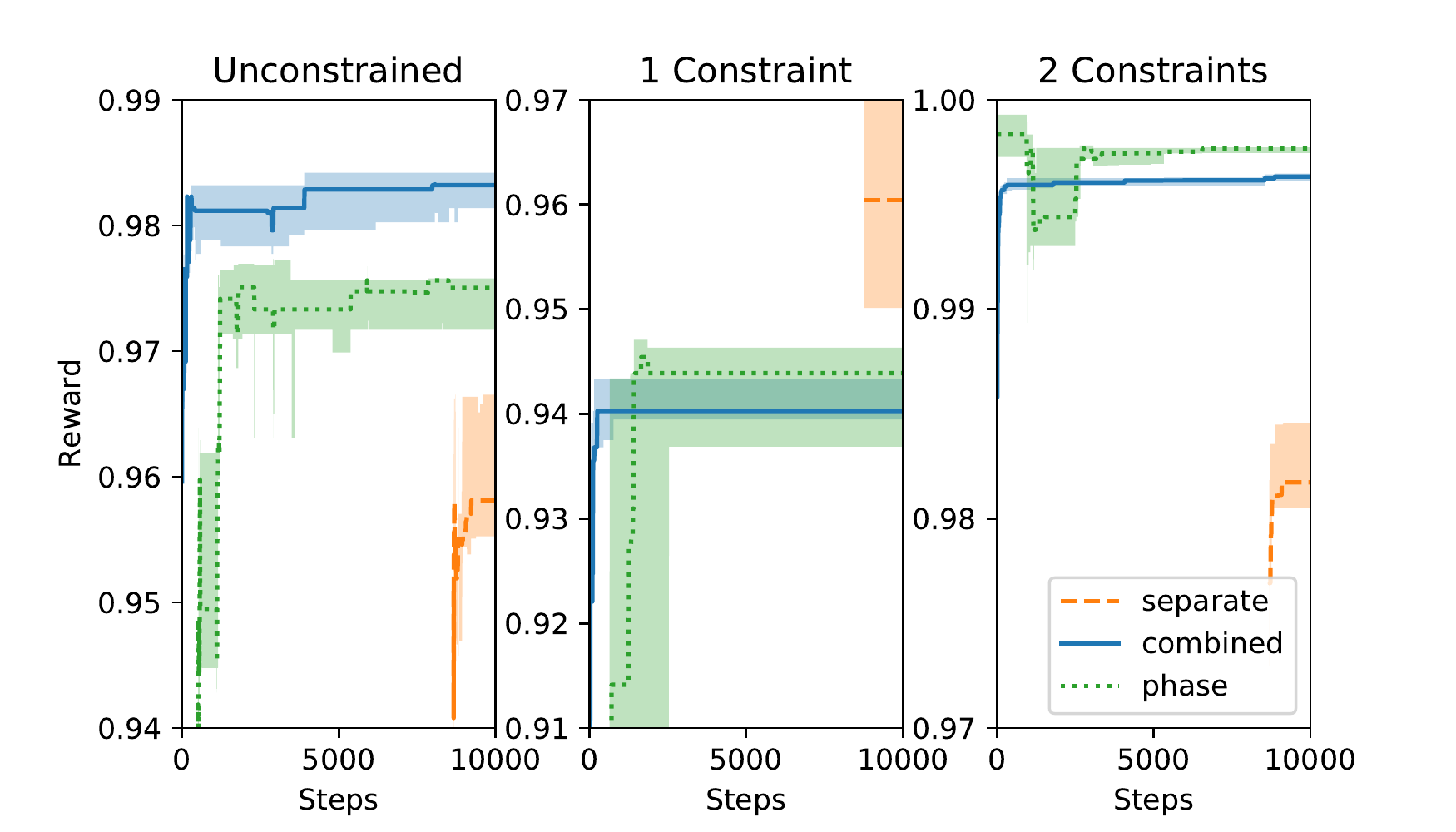}{trim = 0.6cm 0mm 0 4mm}{Reward values of separate, combined and phase strategies for our 3 search scenarios. Note that separate search only has a MOO objective in its second part when it is optimizing the HW accelerator for the discovered CNN.}

As Fig~\ref{pareto_arch} shows, there is a three-way tradeoff between area, latency and accuracy -- to improve a metric, one or both of the other two must degrade.
The Pareto-optimal points form concentric accuracy-latency tradeoff curves, each at a different accelerator area -- different points on the y-axis represent different CNNs, and different points on the x-axis are different accelerator designs. 
%By modifying the CNN, we roughly move along the concentric accuracy-latency curves.
%By changing the accelerator hardware, we move across a horizontal line (thus affecting both latency and area).
Fig.~\ref{pareto_arch} highlights the diversity and number of \textit{good} points, and motivates why automated codesign techniques are necessary.
First, less than 0.0001\% of points in the search space were actually Pareto-optimal -- it is near impossible to manually pin-point these model-accelerator pairs.
Second, the Pareto-optimal points are very diverse and include 338 accelerator variants and 136 different CNN cells -- it is very difficult to manually translate from accuracy/efficiency requirements to an optimal point.
In the next two subsections we present and evaluate NAS search strategies that find codesign points close to these Pareto-optimal solutions.

%-------------------------
\subsection{Search Strategies}
%-------------------------

%To navigate the codesign search space, we explore three alternative search methods.

\subsubsection{Combined Search}
\label{sec_combined}
The first search strategy is to consider both sub-search spaces together as in Equation~\ref{opt_search} and apply REINFORCE directly -- we call this \textit{combined} search.
This strategy has the ability to update both the CNN and the accelerator in each step, and is therefore able to make faster changes to adapt to the reward function.
However, the \textit{combined} search space is much larger, which may make it more difficult to find the best points.
In this approach we run each experiment for 10,000 steps.

\subsubsection{Phase Search}
We explicitly specify specialized phases during the search by freezing one part of the search space (e.g. a specific accelerator) and only focus on the other (e.g. a CNN design).
We would then select the best found CNN, and switch to the accelerator phase to search for suitable hardware.
The two phases are interleaved and repeated multiple times in order to find a globally optimal solution.
This requires us to have two different controllers -- one which only learns to select the best combination of options for the FPGA design and the other one to optimize the CNN structure.
This divide-and-conquer technique may make it easier to find better locally-optimal points (per search space).
However, mutual impact between the phases is limited, which may make it more difficult to adapt the CNN and the accelerator to each other optimally.
When running phase search, we set number of steps for each CNN phase to 1000, and 200 steps for each HW phase, repeating them until we hit the total of 10,000 steps.

\subsubsection{Separate Search (baseline)}
We compare our proposed codesign search strategies above to a baseline where we separately search for a CNN, followed by design-space exploration of accelerator parameters.
This methodology is similar to the conventional, sequential design of the two parts.
We run the \textit{separate} experiments for the total of 10,000 steps splitting the two phases into 8,333 and 1,667 steps respectively.

%-------------------------
\subsection{Search Results}
%-------------------------

\noindent We evaluate our search strategies using three experiments:

\begin{enumerate}
    \item \textbf{Unconstrained}: Zero constraints and we arbitrarily\footnote{The choice of weights is critical in determining the neighbourhood of good points explored, but we do not study that in this work. We refer the reader to prior literature for an in-depth analysis~\cite{moo_book} and a relevant case study~\cite{mnasnet}.} choose the MOO weights as follows $w(area,lat,acc)=(0.1,0.8,0.1)$. This scenario may be useful to simply search for many \textit{good} points to understand the codesign search space.
    \item \textbf{1 Constraint}: One constraint on latency $lat<100 ms$ and $w(area,lat,acc)=(0.1,0,0.9)$. This scenario is similar to when an end-user may know the task and real-time requirements, but is not sure which FPGA device to choose -- the best accuracy at each device size may aid such decision. 
    \item \textbf{2 Constraints}: Two constraints on area and accuracy $acc>0.92, area<100 mm^2$ and we optimize latency (single objective). This occurs when there is a maximum FPGA area budget and a minimum tolerated accuracy for an application -- this is a common use-case. 
\end{enumerate}

Fig.~\ref{searches} plots the top 100 Pareto-optimal points that maximize each experiment's reward.
Therefore, a good search algorithm would produce results in the vicinity of these top Pareto-optimal points.
We also plot the top result from our 3 search strategies, and we repeat each experiment 10 times.
Therefore, we have a maximum of 10 points per search strategy for each plot.
Fig.~\ref{losses} shows the reward function (averaged over 10 experiments) for each experiment.
Note that we only plot the reward function $\mathcal{R}$ and we do not display the punishment function $\mathcal{R}_v$ on the plot. 
In summary, the results show the following trends.

\subsubsection{Separate Search}
``Separate" search cannot consistently find good points within the constraints. 
This is because it searches for the most accurate CNN model without any context of the HW target platform. 
Fig~\ref{e1} shows two \textit{lucky} ``separate" points that are superior to other searches (and that is also reflected by a higher reward in Fig.~\ref{losses}). 
However, the plots do not show that the 8 remaining points all have latencies that are much higher than the constraint. 
This is true for all plots in Fig.~\ref{searches}, where only a few ``separate'' points fit within the displayed axes, while the rest of the points are generally high accuracy but very low efficiency. 
This shows the \textit{randomness} of CNNs that are designed without HW context -- they may or may not fall within efficiency constraints based on chance, even if the accelerator is heavily tuned for the separately-found CNN, further motivating the need of joint codesign.

\subsubsection{Combined and Phase Searches}
These two search strategies improve upon separate search as they take the HW accelerator into account, and more importantly, they consider all variants of the hardware accelerator and all variants of the CNN simultaneously. 
Fig.~\ref{losses} shows that ``combined" is generally better in the Unconstrained experiment, whereas ``phase" achieves a higher reward with both the constrained experiments. 
This is also highlighted in Fig.~\ref{e2} that clearly shows that phase gets closer to the ideal points. 
However, the same figure shows a shortcoming of ``phase" search.  
It is more prone to missing the specified constraints, likely because there are only limited opportunities to switch from the CNN phase to the FPGA phase within the 10,000 steps in our experiment -- if we increase the number of search steps, we expect these two experiments to also find points within the constraints.
More generally, we can say that phase search is slower to converge, compared to combined search. 
This is also highlighted in Fig.~\ref{losses} where phase search goes through a few exploration phases before finding its best result.
In summary, we believe that both of these search techniques have their merits; combined works better when the search is unconstrained and is generally faster to converge to a solution, while phase finds better points when there are constraints but typically requires more search steps to do so.

%=================================================================================
\section{CIFAR-100 CNN-Accelerator Codesign}
\label{sec_imagenet}
%=================================================================================

In this section we use Codesign-NAS to discover a CNN model-accelerator pair that optimizes the task of CIFAR-100 image classification.
We show that Codesign-NAS can exceed both the efficiency and accuracy of well-known CNN architectures even when paired with their optimal accelerators.

%----------------------------------
\subsection{Experimental Setup}
%----------------------------------

Unlike our use of NASBench in previous sections, we have no precomputed results for CIFAR-100 image classification, so we must train all discovered CNNs from scratch.
However, we still use the same codesign search space as defined in Section~\ref{method} to be able to easily reuse our reliable FPGA latency and area models which were verified on our CNN search space.
We use the same training parameters shown in previous work~\cite{nasbench101} with 108 epochs of training, standard data augmentation (padding, random crop and flipping), initial learning rate of 0.1 with cosine decay and weights decay of $10^{-4}$.
Training every new CNN takes approximately 1-GPU hour, so to be able to train many models, we parallelize Codesign-NAS over 6 machines, each with 8 Nvidia-1080 GPUs each, allowing us to train 48 models in parallel.

We run Codesign-NAS with two constraints combined into one.
Specifically, we combine latency and area into performance-per-area ($perf/area$) and constrain the $perf/area$ to a threshold value. We then attempt to maximize accuracy under the constraint.
For our RL controller, we gradually increase the $perf/area$ threshold according to $(2,8,16,30,40)$ -- we run the search for \til2300 valid points in total, starting with 300 points at the first threshold value and increasing to 1000 points at the last threshold value.
We found that this gradual increase in threshold makes it easier for the RL controller to learn the structure of high-accuracy CNNs.
We decided to use the ``combined" search strategy described in Section~\ref{sec_combined} as it has shown to be faster to converge.

\comment{
Before delving into results, we would like to highlight that CIFAR-100 image classification is almost as \textit{difficult} as ImageNet classification, reflected by the fact that its Top-1 accuracy numbers are typically similar to ImageNet~\cite{CIFAR}.
However, CIFAR-100 has a much smaller training set (60K vs 1M), consequently, training a CNN to perform CIFAR-100 classification is approximately two orders of magnitude faster than ImageNet, making it more feasible for the compute infrastructure available for our experiments.
}
%----------------------------------
\subsection{Results}
%----------------------------------

\comment{
\begin{itemize}
    \item Discuss that fair comparison is googlenet and resnet ``cell" not the full network as they have different skeletons (?) let's run first and compare anyways to see where we are.
    \item We give goog/resnet their \textbf{ideal} accelerator -- stress that.
    \item Ideally also add comparison to resnet-50 and googlenet v1.
    \item Basically just point to table and graph and mention key numbers and how much we win by.
    \item Also point to discovered cells
\end{itemize}
}

\figvs{1}{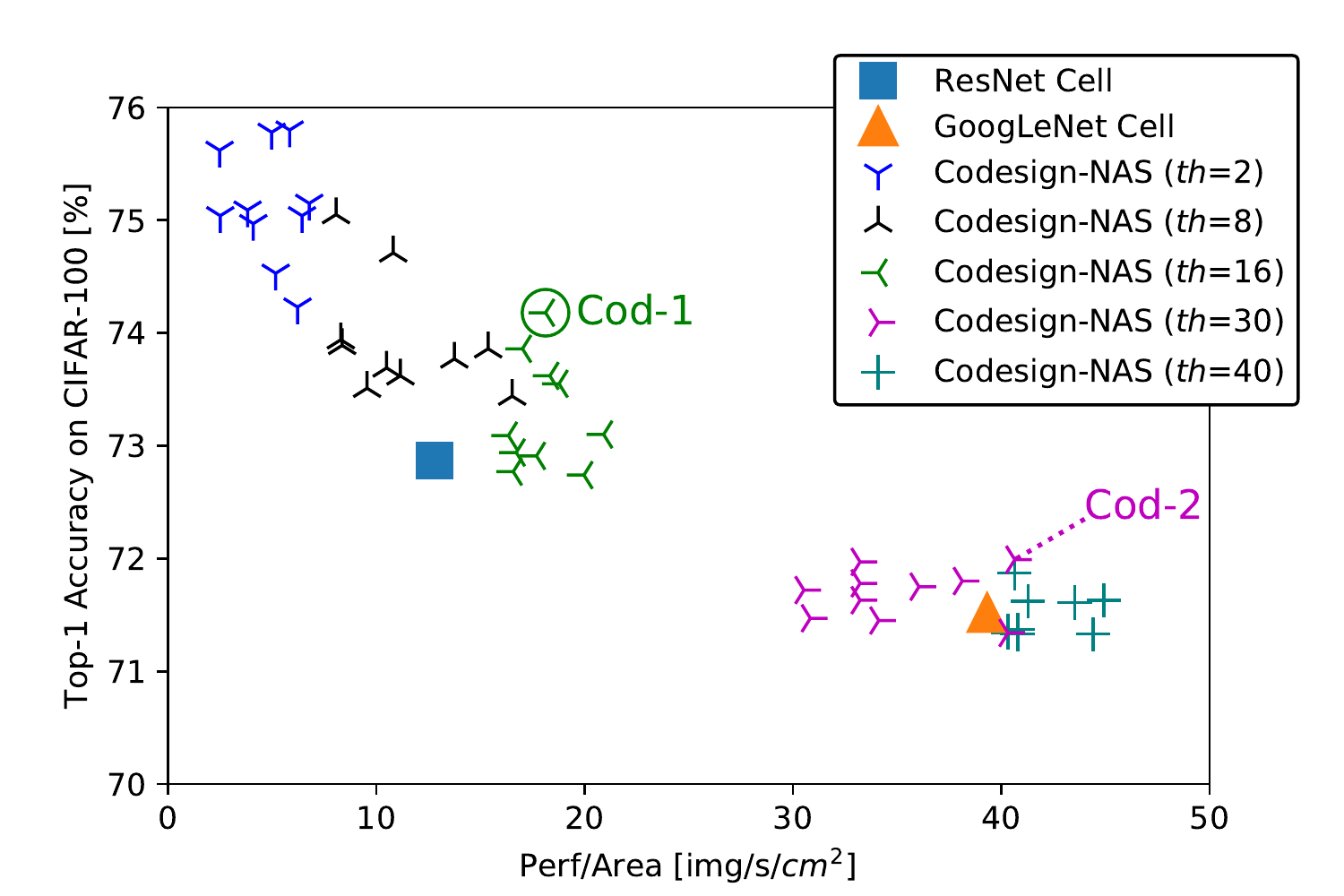}{trim=5mm 2mm 5mm 5mm}{Top 10 points discovered by Codesign-NAS at each threshold value, compared to ResNet and GoogLeNet cells.}

\begin{small}
\begin{table}[t]
    \centering
    \caption{Best points found by Codesign-NAS compared to ResNet and GoogLeNet cells on their best accelerators.}
    \setlength\tabcolsep{3.5pt} % default value: 6pt
    \begin{tabular}{cllll}
         \toprule
         \multirow{2}{*}{CNN} & Accuracy & Perf/Area & Latency& Area \\
             &   [$\%$]  & [$img/s/cm^2$] & [$ms$]  &  [$mm^2$]    \\
         \midrule
         ResNet Cell & 72.9 & 12.8 & 42.0 & 186 \\
         \textbf{Cod-1} & \textbf{74.2} (+1.3\%) & \textbf{18.1} (+41\%) & \textbf{41.8} (-0.5\%) & \textbf{132} (-29\%) \\
         \midrule
         GoogLeNet Cell & 71.5 & 39.3 & 19.3 & \textbf{132} (-0.8\%) \\
         \textbf{Cod-2} & \textbf{72.0} (+0.5\%) & \textbf{40.6} (+3.3\%) & \textbf{18.5} (-4.2\%) & 133 \\
         \bottomrule

    \end{tabular}
    \label{best_pts}
\end{table}
\end{small}

Fig.~\ref{imagenet1} shows the top-1 accuracy and $perf/area$ of various points searched by Codesign-NAS. 
We plot the top 10 points among the model-accelerator pairs visited at each threshold value.
The plot also shows the ResNet and GoogLeNet cells within our CNN skeleton\footnote{We believe that the fairest comparison to discovered cells from Codesign-NAS is to compare against GoogLeNet and ResNet cells within our same skeleton. Alternatively, we can also use our discovered cells within the ResNet-50~\cite{resnet} or GoogLeNet v1~\cite{googlenet} skeletons, and we anticipate very similar findings.} (Fig.~\ref{nasbench_ss}), and we pair those with their most optimal accelerator in terms of $perf/area$.
This is a difficult baseline to beat as we are comparing against two well-known high-accuracy CNN cells when implemented on their best possible corresponding accelerator in our FPGA search space.
However, as the plot shows, we find many points that exceed both the accuracy and efficiency of both the ResNet and GoogLeNet baselines.

We highlight the best two among those points, and label them ``Cod-1" and ``Cod-2" on Fig.~\ref{imagenet1}.
Cod-1 improves upon ResNet by 1.3\% accuracy while simultaneously improving $perf/area$ by 41\% -- considerable gains on both accuracy and efficiency.
Cod-2 shows more modest improvements over GoogLeNet as shown in Table~\ref{best_pts}, but still beats it on both efficiency and accuracy while running 4.2\% faster in terms of absolute latency.

\begin{table}[t]
    \centering
    \caption{HW of best points found by Codesign-NAS.}
    \begin{tabular}{lcc}
         \toprule
          HW Parameter           & Cod-1 & Cod-2            \\
         \midrule
          filter\_par, pixel\_par  & (16, 64)   & (16, 64)               \\
          buffer\_depths   & (4K, 2K, 4K) & (8K, 2K, 2K) \\
          mem\_interface\_width  & 256  & 512                \\
          pool\_en               & false & false             \\
          ratio\_conv\_engines   & 0.33& 0.25   \\
         \bottomrule
    \end{tabular}
    \label{codhw}
\end{table}

\begin{figure}[t]
\centering
\subfloat[Cod-1]{
   \includegraphics[width=0.42\columnwidth,keepaspectratio, trim=0cm 1cm 0cm 2cm]{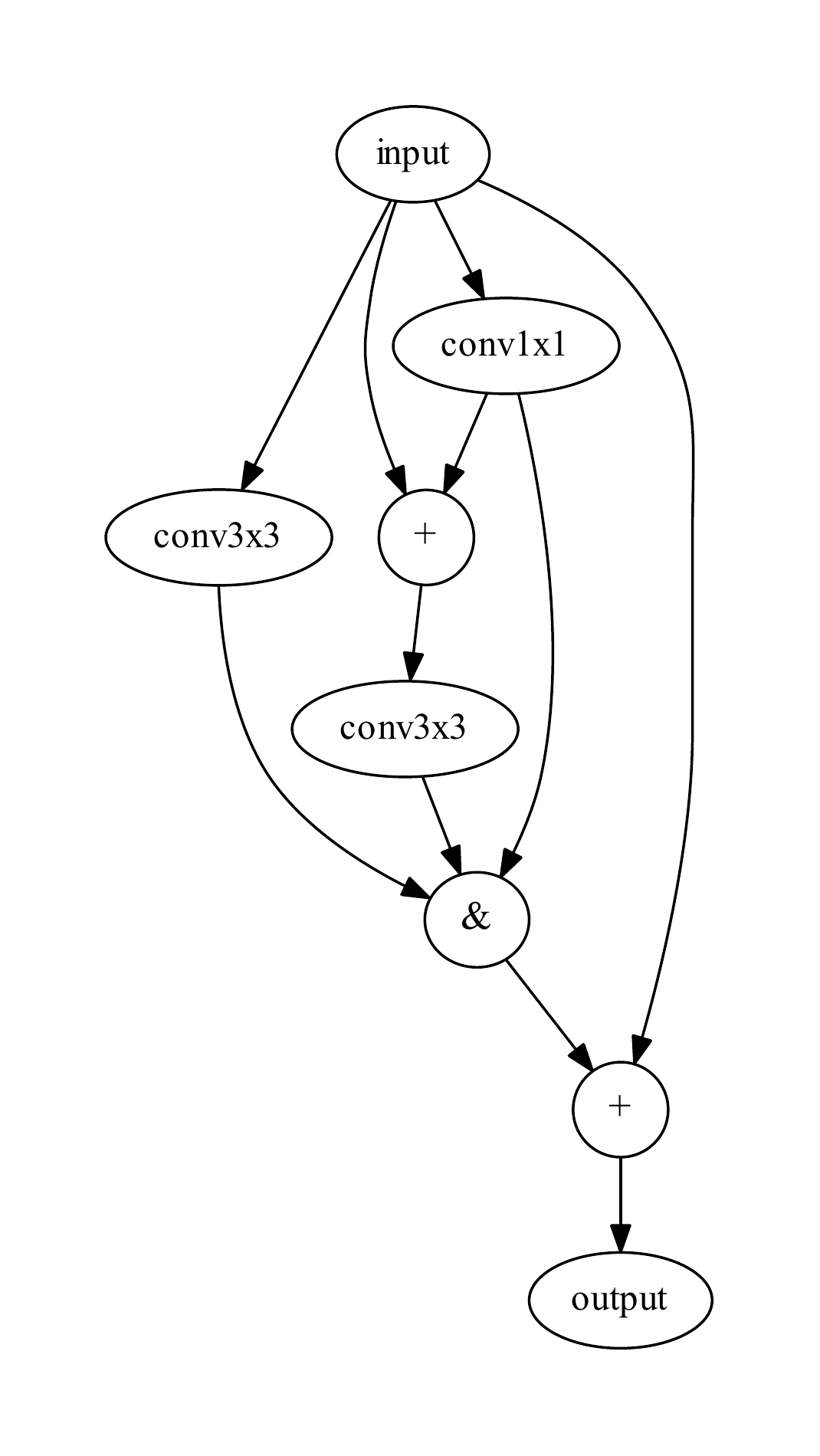}
   \label{cod1}
 }
\subfloat[Cod-2]{
   \includegraphics[width=0.42\columnwidth,keepaspectratio, trim=0 0cm 0 1cm]{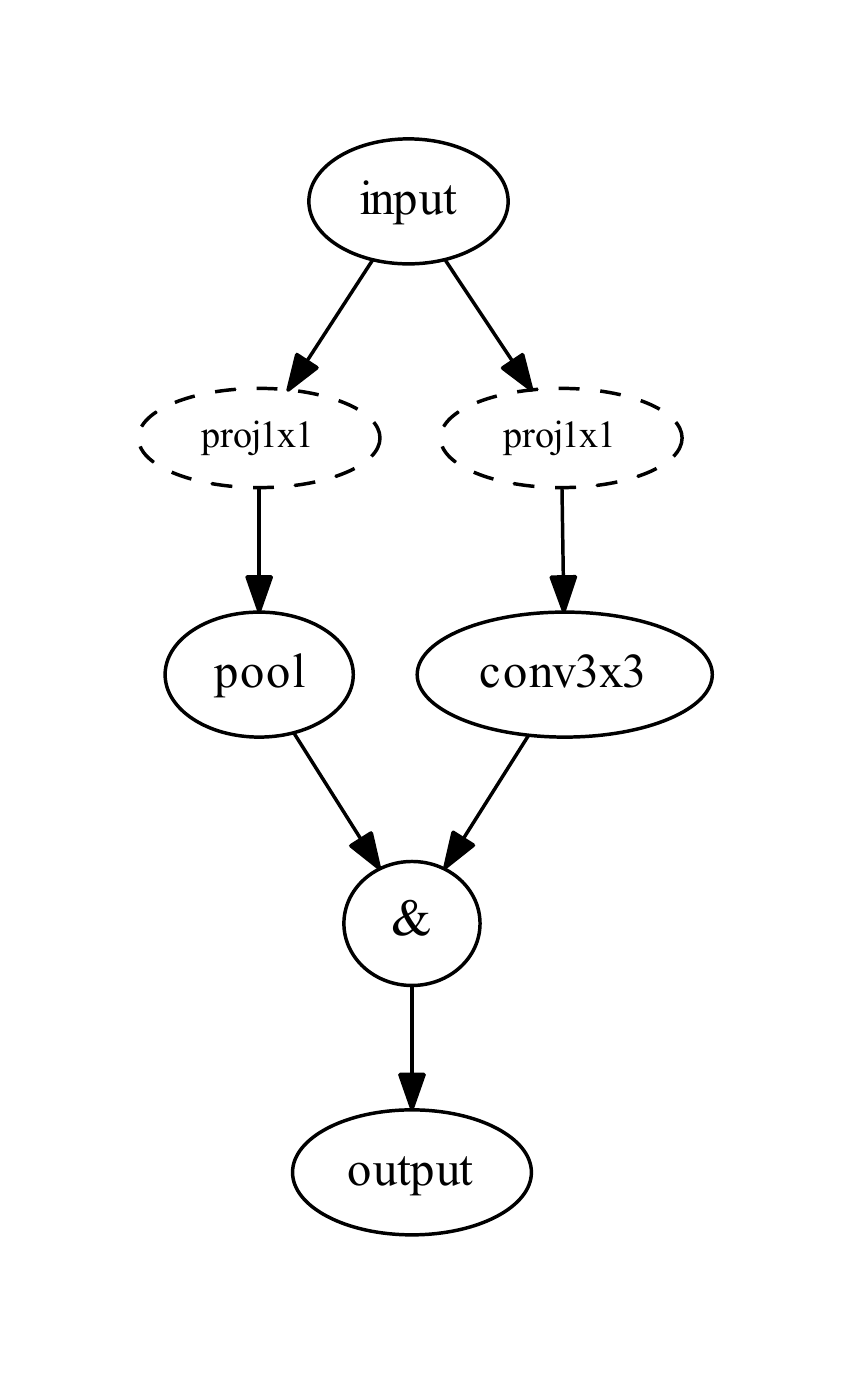}
   \label{cod2}
 }
\caption{Best CNN cells discovered by Codesign-NAS that improve upon (a) ResNet and (b) GoogLeNet cells.}
\label{cods}
\end{figure}

Fig.~\ref{cods} shows the CNN cell structure of Cod-1 and Cod-2, and Table~\ref{codhw} lists the HW parameters.
It is difficult to reason about automatically designed CNNs~\cite{nas_original}, but we will highlight observations of our codesigned model-accelerator pairs.
For example, the Cod-1 CNN manages to beat ResNet accuracy but uses an important ResNet feature: skip connections and element-wise addition as shown by the rightmost branch of the cell in Fig.~\ref{cod1}.
On the hardware side, both Cod-1 and Cod-2 use the largest convolution engine and avoid the use of a dedicated pooling engine. 
However, the other HW parameters are tailored for each CNN.
For example, both the input buffer size and the memory interface width are smaller for Cod-1, likely due to the fact that the Cod-1 CNN uses a larger number of smaller convolutions compared to Cod-2.

Naturally, we anticipate that there might even be better points within our search space that has \til3.7 billion points in total.
We only explore \til1000 points before finding Cod-1 and \til2000 points before finding Cod-2 (compared to 8000 points in prior work to discover a CNN~\cite{mnasnet}).
This highlights the speed of convergence of our RL controller and its effectiveness in finding good designs, especially when properly tuned with representative reward functions and search strategies as we have shown in this paper.

%
%

%=================================================================================
\section{Conclusion}
%=================================================================================

We proposed the automatic codesign of CNN and HW accelerator, and provided a full methodology and case study using FPGAs to support our proposal.
%We introduced Codesign-NAS as a methodology to search for both CNN model and accelerator jointly using reinforcement learning.
We presented three search strategies based on reinforcement learning and compared them against each other and against Pareto-optimal designs.
Finally, we implemented Codesign-NAS for the task of CIFAR-100 image classification on a popular FPGA platform and showed that we can improve upon ResNet (paired with its ideal accelerator) by 1.3\% in accuracy and 41\% efficiency.
These are large improvements, especially considering that our FPGA search space contains a limited set of configurable HW parameters.
We believe that our findings provide a compelling case for the automated codesign of HW and DNNs.
In the future, we hope to study more interesting HW search spaces that give more freedom to Codesign-NAS to tailor a hardware platform for a codesigned DNN.

%###########################################################################################################
%###########################################################################################################

\bibliographystyle{IEEEtran}
\bibliography{IEEEabrv,main}

\end{document}